# How to Analyse Interviews: A Documentary Method of Interpretation

Andy Crabtree, School of Computer Science, University of Nottingham, UK
ORCID ID 0000-0001-5553-6767
Email andy.crabtree@nottingham.ac.uk

**Abstract.** Interviews are commonplace in HCI. This paper presents a documentary method of interpretation (DMI) for the analysis of the topical organisation of talk within a corpus of transcripts, topics that are *'endogenous'* to it and elaborate patterns of social reasoning about issues of relevance to research. The DMI reflexively enables endogenous topic analysis (ETA). We contrast ETA with established qualitative approaches, including qualitative content analysis, grounded theory, interpretative phenomenological analysis, and thematic analysis, to draw out its distinctive character and unique contribution. Unlike established methods, ETA does not require that the analyst be proficient in qualitative analysis, or have knowledge of underlying theories and methods in the social sciences. ETA relies on the DMI, which is a members' method, not a social science method, and mastery of natural language; a competence most people already possess.

## 1. Introduction

Interviews are probably the most common form of study in HCI today. As Hornbæk et al. tell us, "the interview method has helped establish many important findings in HCI and is routinely used in both user research and interactive system evaluation" [40]. This paper introduces a novel method for analysing interviews: the documentary method of interpretation (DMI), which is strongly associated with ethnomethodology [27]. It is, as such, a members' method for analysing interviews that contrasts with established methods in the social sciences, including but not limited to qualitative content analysis, grounded theory, interpretive phenomenological analysis, and of course thematic analysis. These approaches, all commonly used in HCI, turn on 'constructive analysis' [28] and thus treat talk as an object and site for real world theorising [2], employing distinctive coding practices to attach higher-level categories, concepts and themes to interview data. As Saldaña reminds us,

> "A **code** in qualitative inquiry is most often a word or short phrase that **symbolically assigns** a summative, salient, essence-capturing, and/or evocative **attribute** for a portion of language-based or visual data." [72]

The fundamental difference between constructive analysis and endogenous topic analysis lies in *how* codes are developed: constructively, through a general method of interpretation by abstraction, or alternatively, through a members' analysis of the topical coherence of talk. The outcome in both cases is a 'coding frame', i.e., a collection of codes that reflect one's analysis of interview talk. However, one essentially consists of set of higher-level concepts constructed by the analyst to interpret the situated meaning of talk. Whereas the other consists of set of endogenous topics identified by the analyst to inhabit the talk, topics oriented to and elaborated by parties to the talk (interviewer and respondents). The DMI thus enables endogenous topic analysis (ETA) and, in being a members' method, provides researchers with a method in which they are already skilled. No training in methods of constructive analysis (social science) is required. Rather, the analyst needs to hone their pre-existing membership competence and orient to talk's naturally accountable (rather than theoretical) features.

Below we explain the methodological foundations of the DMI and provide a practical example of what is involved in doing it. Taken together, this methodological account and worked example constitute a 'tutorial problem' that **exhibits** the phenomenon it describes – i.e., endogenous topic analysis – and makes it available to others as "instructed action" or "instructions for doing real-world jobs" (ETA in this case) as per ethnomethodology's approach to pedagogy [29]. As Garfinkel reminds us,

> "There is a certain way of reading the text as instructions. If you don't read the text instructionally, but instead you read the text straight-out, say, as a description of a procedure that was done … then you will fail to get

> what is to be done and what is to be learned when the text is read as instructions for doing what the text describes." (ibid.)

What the reader gets out of this paper very much depends on their orientation then. Read it straightforwardly as the retrospective description of a procedure done by way of example and the phenomena will disappear. Read it prospectively as **an instruction for doing**, and what is to be done (analysing talk using the DMI) and what is to be learned (the endogenous topical organisation of talk) will come into view. The reader will come to see how as instructed action the DMI enables the identification and analysis of endogenous topics across a corpus of transcripts to elaborate distinctive patterns of social reasoning of relevance to research. We contrast this novel approach with approaches commonly used in HCI to analyse interviews, paying particular attention to how established approaches work and do the business of coding, to make the fundamental difference and novel contribution of this paper patently clear.

## 2. The documentary method of interpretation: a tutorial problem (part 1)

We have said that the DMI is a members' method. We briefly explain what we mean when we talk about members' methods in general before we outline the DMI. We also explain what a member is and outline some salient methodological features of talk for members, which are relevant to using the DMI to analyse interviews.

### 2.1 Members' methods and the DMI

When we talk about members' methods, we are talking about the unique focus of ethnomethodology. Doug Weider explains our distinctive interest:

> "Garfinkel coined the term 'ethnomethodology' … after seeing interests cognate to his own in the developing 'ethnosciences' … The initiating idea of the ethnosciences was the notion that the knowledge possessed by members … could be viewed as analogous to the knowledge systems of the sciences … Along the same lines, one could also propose an ethnosociology which inquires into folk theories, concepts, methods of theorizing, and the like … The word 'methods' in ethnomethodology means simply 'methodology', as in the methods of science ... Just as we have found analogues to botany in the field of ethnobotany, we can find analogues to scientific reasoning in the reasoning of 'the folk' – hence, we have ethnomethodology … since the sciences are also social activities, they too are analyzable as ethno- or folk-disciplines, so eventually … ethnomethodology would also describe and analyze the methods of the sciences as well." [81]

The DMI is an ethno-method, a members' method, of fact-finding. Garfinkel attributes the DMI to Karl Mannheim and draws on it to make the problematic status of social science methods of inquiry visible, including supposedly rigorous quantitative methods. Garfinkel recognised that quantitative research relies on qualitative interpretation (see [17]) and that the act of interpretation itself depends on the documentary method.

> "The method consists of treating an actual appearance as 'the document of', as 'pointing to', as 'standing on behalf of' a presupposed underlying pattern. Not only is the underlying pattern derived from its individual documentary evidences, but the individual documentary evidences, in their turn, are interpreted on the basis of 'what is known' about the underlying pattern. Each is used to elaborate the other." [27]

Garfinkel used the DMI to furnish a demonstration of ethnomethodology's claim that the sciences are analysable as folk disciplines [31]. The DMI is essentially **a method of pattern recognition**. Indeed it is a great many methods of pattern recognition, a constituent feature of fact-finding in all manner of circumstances, having its own local or situated character [77]. The kind of patterns we find in using the DMI to analyse interviews consist of **discrete orders of practical reasoning** elaborated by the **endogenous topics** built-in to members' talk. Endogenous simply means that topics of talk are *internal to* and *originate within* members' conversation, that they are an *inherent* feature of whatever it is members are talking about.

### 2.2 Member

As Garfinkel and Sacks [32] tell us, the notion of member stands at the heart of the matter. They go on to say,

“We do not use the term ‘member’ to refer to a person. It refers instead to the **mastery of natural language**, which we understand in the following way. We offer the observation that persons, in that they are heard to be speaking a natural language, *somehow* are heard to be engaged in the objective production and objective display of commonsense knowledge of everyday activities as observable reportable phenomena. We ask what it is about natural language that permits speakers and auditors to hear, and in other ways to witness, the objective production and objective display of commonsense knowledge, and of practical circumstances, practical actions, and practical sociological reasoning as well? What is it about natural language that makes these phenomena observable-reportable, i.e., *account-able* phenomena? For speakers and auditors the practices of natural language somehow exhibit these phenomena in the particulars of speaking, and *that* these phenomena are exhibited is itself, and thereby, made exhibitable in further description, remark, question, and in other ways for the telling.”

A member has mastery of natural language. It could be English, Chinese, Arabic, Urdu, etc. It is whatever language we learn to speak in everyday life, which we share in common and practice with those around us. Mastery of natural language does not denote a special or advanced qualification. Of course there are people who cannot speak a natural language or who are still learning, but to have mastery of natural language simply means one can engage in mundane discourse and understand what those around us say in all manner of practical circumstances. Practical circumstances are crucial. As Garfinkel and Sacks point out, in speaking a natural language, persons are *somehow* heard to be engaged in the objective production and objective display of commonsense knowledge of everyday activities as observable reportable phenomena. Having mastery of natural language is not simply a matter of being able to talk then. Rather it reflects the fact that natural language is bound to everyday activities, to practical action and practical reasoning in a diverse array of practical circumstances that members possess commonsense knowledge of, it is ‘indexical’ to use an ethnomethodological phrase (ibid.). What is important about being a member then, is that it provides the **necessary competence** with which to understand and analyse the real-word character of talk as it naturally occurs in everyday life and everyday situations, including interviews. Membership competence is a competence that most adults possess, acquired through our engagement with society, not special academic training.

### 2.3 The real-world character of talk: some salient members’ methods for organising and analysing talk

Harvey Sacks’s *Lectures on Conversation* [66] provide a wealth of studies of members’ mastery of natural language and their methods for organising and analysing talk. Before we outline some salient methods for analysing interviews, let me offer an observation: members can often read and make sense of the talk contained in interview transcripts. This isn’t to say they understand everything that’s said, there are limits as we’ll discuss in due course. But for the most part, insofar as interviews are conducted in a natural language members have mastery of, transcripts are hearably replete with mundane meaning. Hearable doesn’t mean audible. It means that members – masters of natural language – **understand** what persons are saying and that they are able to follow their talk as it unfolds. How is that possible? Well one thing that happens when members listen to other members speak is that they hear not only words, but words strung together in **intelligible utterances** that in being spoken articulate **mundane reasoning** [64]. Thus, and for example, we hear what participants in an interview think about this or that, how they experience it, what they make of it, what they find problematic about it, what they like or don’t like about it, etc. Furthermore, when we read a transcript and listen to participants’ (members’) mundane reasoning we can hear that it is “**heavily topically oriented**” [69]. A topic is not a technical construct, it’s not akin to a category or theme introduced by the analyst. Topics are ‘**oriented objects**’ [30], i.e., objects members orient themselves to in conducting conversation, the major constituent of talk as my dictionary puts it, and thus **a primary organisational feature of talk**. As Boden and Bielby [6] tell us, topic is the interactional stuff of conversation. Topics are what conversationalists (members) – including interviewers and respondents – orient themselves to and busy themselves with in their talk together.

Given that topics are key to members’ talk, we should pay careful attention to them. When we read a transcript we can hear topics being introduced, often, at first, by the interviewer. We can see as Harvey Sacks’ studies show [69] that conversationalists are very respectful of topics, that they exhibit their attention to talk by speaking to the topics raised, and that they preserve the conversational salience of topics through turns at talk. We can see, as Sacks makes perspicuous, that there is a "stepwise movement" of topics, that topics flow from one to another, and not simply because the interviewer asks a different question but because topics "can also be used to make jumps" to new topics [71]. Thus topics shift and even change over the course of discussion [69]. As Sacks tells us, the dynamic nature of "talking topically" means "you can't be assured that the topic you intended is the topic they will talk to" [70]. Participants appropriate topics, make them into something they can speak about, particularly through the introduction of "**sub-topicals**" [71]. Again, sub-topicals are not objects introduced by the analyst, they are a constituent feature of members' talk that **elaborate topics**. As Sacks notes, "talking topically doesn't consist of blocks of talk about 'a topic' [it] constitutes ways of talking which involves attention to '**topical coherence'** [70]." This is to say that members don't engage in extended sequences of talk about a topic. They talk about things that **relate** to the topic, i.e., sub-topicals. Members' talk thus has topical coherence, which is furnished through an array of sub-topical utterances that relate to and elaborate the topic from their perspective.

Topical coherence provides members with their interpretative frame. It is the taken for granted background against which talk is understood by members. A common background expectancy we have as masters of natural language that talk *will* be topically coherent no matter its particulars, used and relied upon to interpret participant's utterances in the ongoing course of conversation. The mundane work of interpretation notably entails hearers attending to talk to see if the sub-topical remarks offered fit in with and preserve the orientation to a present topic, or if they mark shifts in and changes of topic [69]. This practical accomplishment relies on the **hearer's maxim**, which Sacks defines in saying "**if there are two categories used, which can be found to be part of the same collection, hear them as part of the same collection**" [67]. More generally we can that say if two sub-topical utterances can be heard to be speaking to the same topic, then treat them as such. The hearers' maxim enables members to make determinations of topical equivalence or divergence within a sequence of talk. It can be used alongside the documentary method to also make determinations of topical equivalence or divergence **across a corpus of transcripts**.

### 2.4 How to analyse interviews

An ethnomethodological approach would have the analyst be in a "concerted competence of methods" with members [34]. We should not try to ignore or suspend our membership competence then, our mastery of natural language provides us with the fundamental skills we need to analyse the real-word character of talk, no special training is involved or required. Rather, we should be sensitive to the methodological character of real-word talk as outlined above. Thus, talk for members consists of intelligible utterances that articulate mundane reasoning. Mundane reasoning is topically oriented. Talking topically implicates an array of sub-topical utterances that cohere and elaborate topics. Members use the hearer's maxim to determine if sub-topicals are equivalent or divergent, i.e., continue the current topic or change topic and introduce a new one. Sacks tells us we should take the collection of sub-topicals that cohere topically and treat it "as a unit" [67]. We will call this unit an **endogenous topic** to reflect its origin in talk. An endogenous topic is a members' topic. It consists of collection of sub-topical utterances that have topical coherence and can thus be heard by members to be part of the **same** collection. An endogenous topic need not be restricted to a single transcript. It is a methodological feature of interviews that they discuss the same topics with members. We can look for endogenous topics across a corpus of transcripts then, search for an array of sub-topicals that elaborate them. The search is driven by the hearer's maxim tied to the DMI. It allows the analyst to treat an actual appearance (specific utterances) as 'the document of', as 'pointing to', as 'standing on behalf of' a presupposed underlying pattern (an endogenous topic). Not only is the underlying pattern derived from its individual documentary evidences, but the individual documentary evidences, in their turn, are interpreted on the basis of 'what is known' about the underlying pattern. Each is used to elaborate the other. Thus, as we start to collect sub-topical utterances together through the use of the hearer's maxim, the emerging pattern, the endogenous topic, becomes a complimentary resource for deciding if other sub-topical utterances are equivalent.

**3. Doing endogenous topic analysis (ETA): a tutorial problem (part 2)**

The corpus of transcripts used in the following example are publicly available: http://doi.org/10.17639/nott.7408. Part 1 of our tutorial problem provides an ethnomethodological orientation to the analysis of interviews. Part 2 of the tutorial problem provides an example of using the documentary method of interpretation to do endogenous topic analysis. Together Parts 1&2 of our tutorial problem constitute what Garfinkel calls a "lebenswelt pair" [30]. Lebenswelt pairs are distinctive pedagogical resources for teaching ethnomethodology. They are composed of constituent segments. A first segment of the pair is a "classic account" that "speaks on behalf of practical action." The second segment is a description of "the way of working" involved in accomplishing practical action. Garfinkel tells us, we cannot teach **the way of working** if we only address the first segment, we must "recover the pair." He tells us that the pair cannot be read off the page, no matter how talented the reader; that "work-site practices" are not equivalent to procedures specified in classic accounts, and that they are unavailable to the "situationally disengaged." They are only available to practitioners as "embodied equipmentally affiliated skills." Praxiological description, i.e., description of the embodied equipmentally affiliated skills constitutive of work-site practices, should be read as "instructions for doing a real-world job"; endogenous topic analysis in this case.

Our example is taken from a study already reported in a peer-reviewed journal [24], so we will not labour the details of the case any more than is necessary to furnish the second segment of the pair. The study was of a design fiction configured as a breaching experiment. It was designed to elicit taken for granted and thus usually unspoken expectations members have of a commonplace social phenomenon: how watching TV works. The design fiction allowed participants to get hands-on experience of 'smart adaptive TV', where media content is adapted in real time by artificial intelligence exploiting smart home devices to offer viewers a personalised and immersive experience. We wanted to know what people would think about the use of artificial intelligence to create bespoke media experiences, particularly if it met their expectations about how watching TV works, and should work in the future, and if it might therefore be something they want to adopt in their everyday lives. This is what motivated the interviews in our example, why the data was collected, the problem we sought to address.

A variety of interview formats could have been employed to help us address the problem: structured or semi-structured approaches are often used; a looser, more open approach could be adopted to drive exploration; or a data-driven approach might have been used as one sometimes finds in HCI [e.g., 78]. In this case, we wanted the interviews to be open and responsive to participant's experience and viewpoints, so we seeded them with a set of basic exploratory questions worked up by 3 of us at a whiteboard. But it doesn't really matter what the format is: talk is talk and is as such available to membership competence regardless of the interview format.

Our work at the whiteboard yielded the following categories and list of questions:

**a) Reflections on the experience**

How did it go, what do you think?
What did you like?
What didn't you like?
Did you understand AI was controlling the experience?
How do you feel about that?

**b) Reasoning about AI**

How would you feel about having this smart technology and AI in your home?
Would you let AI access your personal devices to deliver other kinds of engaging experiences?
Would you let AI access your data to create a profile it can share with other parties so they can deliver personalised services to you?
Would you let AI control other aspects of your everyday life?

**c) Reasoning about control**

Would you like to be able to control what devices AI can access?
Would you like to control what AI does with your data?
Would you like to control who AI shares your data with?
Is human control always needed for you to trust AI?

At the bottom of the list it was written: *"Explore mundane reasoning implicated in responses: why, why, why? Open up reasoning with participants in making their choices and decisions."*

The smart adaptative media experience was installed in a mobile platform, a bespoke caravan [21], and the interviews took place at two different locations: at the University of Salford's Media City building, Media City is a major centre of TV production in the UK, located at Salford Quays in the North of England; and the Bath Digital Festival, a community event in the south-west of the country that raises awareness of new technologies and provides a platform for community members to build connections locally and nationally. Two very different venues then. We positioned the caravan close to a café at both sites to attract passers-by. The caravan piqued curiosity as we hoped it would do and a broad range of participants was subsequently recruited from the flow of people into and out of the cafés, from those who work in media production through to technology developers, students, university staff, event attendees and members of the public at large who were having a day out at Media City or the Digital Festival. Consent to participant in our research was obtained at both venues, and took around 10 minutes to complete. Participants then entered the caravan and took part in the smart adaptive media experience, which lasted around 20 minutes. The interviews were conducted after the experience and involved 30 participants over 3 days at Media City, and 20 participants over 2 days at Bath. 50 participants in total. The interviews lasted on average around 15 minutes each and were recorded on audio and subsequently transcribed automatically. We could have transcribed the data differently, e.g., using the conventions of conversation analysis [42], but we aren't trying to do the kind of analytic work that conversation analysts do. For the purposes of analysing interviews, a practically adequate representation of what was said will suffice, where adequate means sufficient to reflect mundane reasoning. Transcripts generated automatically will do the job, insofar as their mistakes are repaired either before or during analysis such that we can discern members' reasoning and the topical coherence of their talk. The corpus of transcripts furnished 178 pages of data totalling 67,000 words.

### 3.1 The embodied equipmentally affiliated skills of endogenous topic analysis

We have already provided a methodological account describing our analytic approach in section 2 above. Here we provide a praxiological description, i.e., a description of the work practices involved in actually doing endogenous topic analysis (ETA). Remember, an endogenous topic is **a collection of sub-topical utterances** – actual utterances found in multiple transcripts – that **hearably go together**, i.e., have topical coherence. As a matter of work practice, we're looking for sub-topicals. We won't find blocks of talk about a topic when we read transcripts. We'll find utterances that have **topical coherence**, which relate to and speak to topics, and in being pulled together by us in our work with and across multiple transcripts come to form a unit: an endogenous topic. To find them we need to apply the hearer's maxim to identify utterances that hearably go together and coalesce topically.

A good place to start is with the interview questions. They furnish a basic set of topics to start your analysis with, just as they furnished a starting point for the interviews. So start by **mapping responses to questions asked**. Actually try doing it – as Garfinkel tells us, embodied equipmentally affiliated skills are not available to the situationally disengaged reader – I have provided transcripts, so you can give it a go. Go to your computer, download them, open the file, open a Word doc or similar, welcome to the work-site. Try and work with me. Read my description of the work as instructions for doing it. Try and follow by doing it yourself. You can use your own data if you want, but try and do what I am doing.

I am working with the three basic categories of question listed above: questions reflecting on the AI-driven adaptive media experience; questions about having this kind of experience in your own home; and questions about exercising control over AI. Like me, you can start with any transcript (they weren't all bundled together for download when I got them), it really doesn't matter. What matters is that **as you read**, you **pull out discrete extracts of topical talk** about the questions asked, and that you **put them in another document under a top level heading that reflects the questions** (cut and paste) so you can easily navigate the document and continue to add to it. I find the styles function in Word is really useful. You can create and view headings in the sidebar and quickly navigate between them as you drop extracts under them (see Figure 1). Just drop extracts in for starters. Don't try to organise them beyond putting them under the relevant question heading. Don't pull out isolated words or sentences either. Include the question and response, or multiple responses if more than one person is being interviewed, and follow on questions and responses if they continue to elaborate the topic being discussed. You want to retain as best you can the **sense the talk has to the participants**. You don't want to lose that.

Having pulled an extract of talk out of the transcript and placed it in your initial analysis document, **highlight what is topically interesting** about it so you can quickly refer back to it. Just highlight it. You don't need to make notes. You might edit long interchanges. Cut them down a bit to focus on utterances of interest within a stretch of talk. But don't overdo it. Don't lose the sense of what is being discussed and why. For example:

**What did you think about the experience of the caravan?**
P16B: For me it's crazy, because **if I think this is the future, and every machine or engine is speaking, it's a little bit afraid for me.**
P17B: I think it's cool that you have a custom designed living room and all the machines can design or make you feel more comfortable … but it's not on an emotional level, and that's what I don't like. Because **it's like, what am I doing? I'm in front of a machine telling me what I like and what I want to be. I can't relate to that.**
P18B: Yeah, I agree. The most bizarre moment was when the machine was changing the voice and … like responding to our faces … **I was like, what did I do? What did they see on my face to react like this?**

This is all you need to do for starters: cut and paste extracts of topical talk, put them under a question heading, and highlight what is topically interesting about them.

As you work through the transcripts, you will find that some questions are quite one-dimensional, i.e., topical talk follows much the same line of reasoning. Reflecting on the smart adaptive media experience, as above for example, we found participant's reasoning was quite negative. Not only was it found to be something to be a little bit afraid of, something that couldn't be related to, something whose actions where unfathomable, participants similarly found it "unnerving" and "weird" because "it's in control." Participants reported that "The idea that [TV] would be personalised to me, that's not a positive. I'm not interested in that. I'm interested in watching the same show [not] discussing that with someone who has a completely different experience." I don't want to be part of this world going forward, I don't like it, it scares me." "I don't know what to trust." "It did nothing to alleviate my fear of AI." "My assumption now [is] that almost everything is connected to the internet in some way and that everything is connected to a corporation that [has] not necessarily got my best interests at heart."

Nevertheless, responses are often more **topically diverse** and you will find that extracts begin to cohere in manifold ways, such that you end up with different collections of extracts cohering around and elaborating an array of endogenous topics that emerge in response to the interview questions. On the left of Figure 1 we can see a list of endogenous topics found in analysing the interview transcripts on smart adaptive media. The list reflects the categories of questions asked and a collection of topical responses to them. We can see that the questions to do with *having AI like this in your home* (i.e., smart adaptive media) and being able to *control* it have a lot of entries underneath them. That each question is elaborated by a range of topics. Such as being *cautious* or *wary* about adopting smart adaptive media and *not wanting it* in the home, not allowing it to access personal *devices*, or *data*, etc. The reader can see too in the document on the right of the list, that various sub-topical extracts are placed under and elaborate each topic. You can see that the names or labels given to the topics on the left reflect the utterances on the right, sometimes literally (e.g., cautious / anxious / don't want it), sometimes by characterising what the topic is about (e.g., not my devices). You may recall that in introducing the paper we talked about coding. These labels can be construed of as codes, and the collection on the right as a coding frame that reflects our interpretation of the data. As Saldaña told us, a code is a "summative, salient, essence-capturing, and/or evocative attribute for a portion of language" [72]. The attribute in ETA is endogenous, it's visibly, hearably, an **inherent feature of the talk**, either literally or summatively as provided for by a short characterisation that covers a series of utterances. The label, the code, is not an analytic category that I impose on the data based on some extraneous theoretical considerations. It is a descriptive reflection of talk's topical character, as can be seen and heard in the associated utterances.

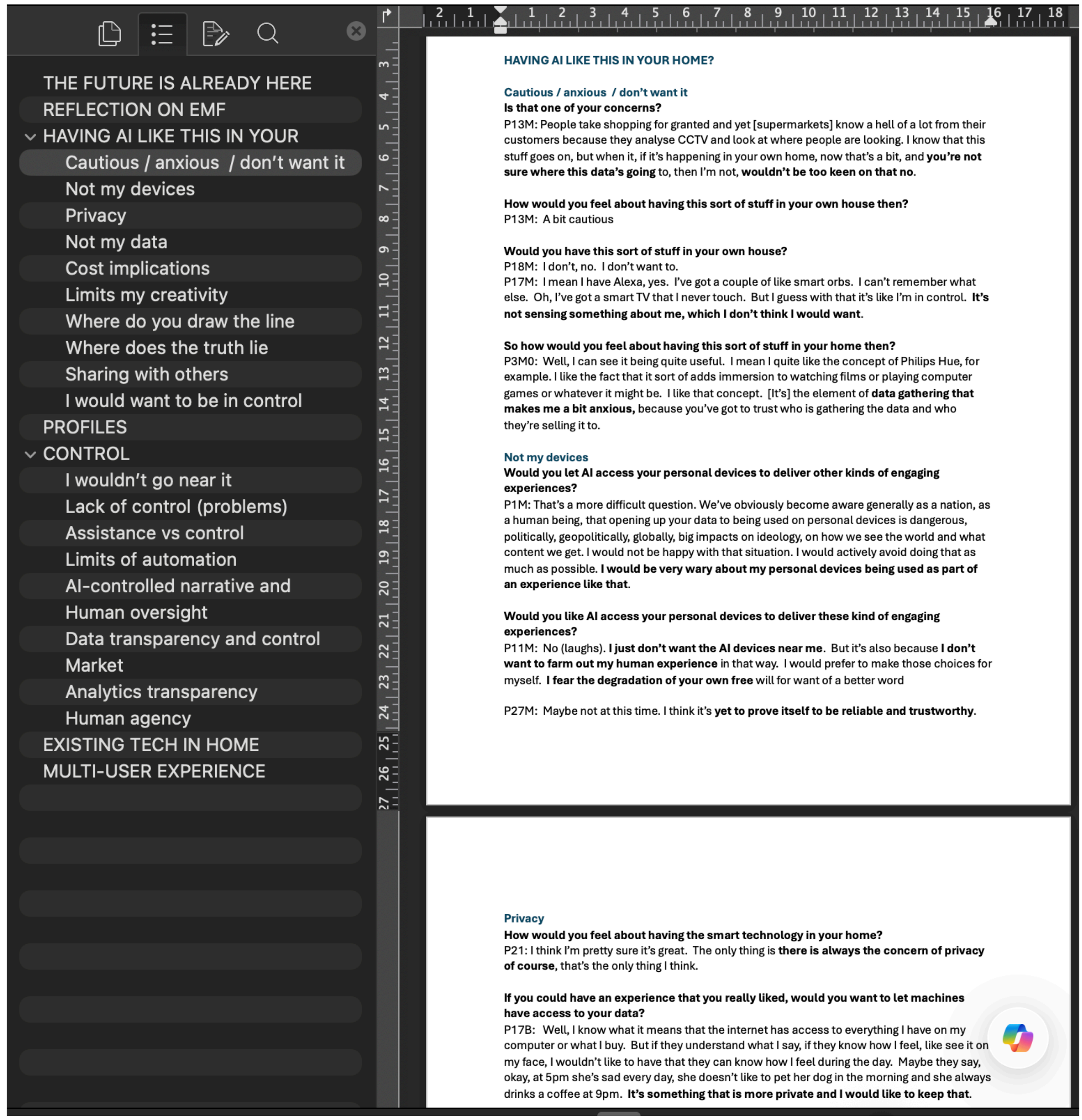

**Figure 1.** Analysing smart adaptive media interviews.

Figure 1 is a screenshot of my actual working document. I use Microsoft Word, there's no reason anyone else has to, but it works for me. It allows me to quickly cut and paste extracts from interview transcripts as I read through them and drop them under question headings; and it allows me to collect extracts that hearably go together under sub-headings that reflect particular topics. Figure 1 reflects the finished article, after a first pass through the data. Were you to look at the document **in flight**, in the course of analysis, it would look considerably **messier and full contingencies**. Extracts wouldn't have coalesced around particular topics, some might be placed together, some might not, and some never will be. There might be a loose working title here and there that reflects what topic is about, e.g., stuff about privacy or data sharing, but there wouldn't necessarily be sub-headings and certainly not a complete set. You'll find that in flight, your working document, your analysis, your interpretation of the data is fluid, in motion. That you collect extracts together on the fly and in doing so come to see that actually something richer is going on. That you have two or three hearably distinct collections of things in hand, not one, for example, and that you have to move things around accordingly. Or that what at first seemed like two different sets of sub-topicals are actually speaking about the same thing, as made perspicuous by some new extract that connects them together. Analysis is full of contingencies and practical choices. The hearer's maxim allows you to resolved them. To determine if these utterances are talking about the same sort of thing as those utterances or not.

Importantly, the situationally engaged reader will also notice if they look at Figure 1 that the list on the left is **not confined to the categories of question asked** and the endogenous topics that elaborate them. Other topics are listed: *multi-user experience*, *existing tech in the home*, *profiles*, and *the future is already here*. As I read through I find that participants discussion go beyond the questions asked of them. That utterances do not easily fit in with them as responses. That they have their own topical character. It is as Sacks told us it would be: "you can't be assured that the topic you intended is the topic they will talk to" [70]. Indeed, what is asked and what we get back in response may be two very different things. Thus, and for example, our participants wondered how smart adaptive media will deal with the mixed requirements of multiple users; they discussed the technology they already have in their everyday lives; they had a lot to say about being profiled; and they made it patently clear that the future is very much here already. These were not necessarily issues that we set out to explore, but again as Sacks told us, topical talk can also be used to make jumps to new topics [71]. This means it is not sufficient to simply map responses to questions as we work our way through the corpus of transcripts. It is also important to **put sub-topicals in focus**, not only the questions.

Take, for example, the following exchange:

How does it make you feel about having these sorts of smart technologies and AI in your home?

P4M: (Sighs) Oh it's difficult. I guess, I guess **I try and forget about it**. I almost exchange the utility of them for some, the understanding that they're going to do something with my data or act in a way that I don't understand. Like you sort of, maybe you buy a new thing and you feel a bit fearful with it for a while and then slowly you forget about it and that's how I feel how I've done with my smart devices. **I've forgotten that they're doing a lot of this stuff.**

P6M: I don't know, well, we all carry – I don't really like the idea that – like especially with the Google Homes and Alexa, it's just a microphone in your house and I don't really like the idea that anyone with, like you know capable, could tap in and listen, because they could. However, I also carry a phone with me everywhere I go which has that capability, so you can't, you know. **It's either all or nothing I guess**.

The question was intended to be prospective in character, about a projected future world in which smart adaptive media resides, but it wasn't treated that way. It was treated as speaking about the world here-and-now, a world in which smart devices already live: "I try and forget about it." "I've forgotten that they're doing this stuff." "It's either all or nothing I guess." So we have, characteristically and by way of example, discovered at random on reading through a transcript two sub-topical utterances about living with smart technology here-and-now. We weren't asking about that, but that's what interested our two participants and if it interests these two it might interest others, so we better pay some attention to it and see where it takes us. Thus utterances to do with living with smart technology here-and-now become **something to look and listen for** as we continue to read through the transcripts, just as utterances to do with any other endogenous topic do. Importantly, **the looking is not constrained by the question**, but by what more participants may have to say about the endogenous topic to hand.

Thus, if we look across the different categories of question and responses to them in the transcripts, we come to find other sub-topicals that complement our two random utterances, utterances that hearably go together (have topical coherence) and further elaborate what they are talking about. For example:

*Reflecting on the experience*

How did it make you feel that there was something taking control, the AI was doing things with your data?

P10M: **I think we just live in a world where our data's passed quite freely by our controlled actions these days anyway**, and we're kind of willing to that. So if we abstract it to the future that's like that's what's going to happen is these devices that we have round our house will consume that data, and I guess we consent to that by plugging them in and kind of using them. I don't know whether we know what we're consenting to. That's my perspective.

*Having this kind of experience in your own home*

How do you feel about giving it [AI] access to your personal devices?

P3M: It already has it. Like everyone's got a smart device pretty much at this point. If you have a modern phone then you have a smart device. If you have, you know, a Fitbit or any kind of wearable, you've got a smart device. **Our data's being collected a million different ways already**. **There's any number of different ways I think AI's being used in different services to like gather your metrics**.

*Controlling AI*
Would you see them as your devices under your control?
P1M: I like to see them as my devices, but **I'm aware that the data is often out of my control.** I see the devices as essentially the **devices that I own**, but I do know that **they're also leaky**.

In terms of control, would you let other services have access to that data from another company?
P5: It's like almost impossible not to nowadays. **I think people have just given up the fight in all honesty**. People still care about it, but there's nothing they can do
P7(2)M: **We're cyborgs already**. **We're augmented humans, aren't we?** How I find my way about, how I plan my day, how I do lots of mundane tasks, even pre-AI, are heavily standing on the benefits of technology.

Asked about sharing data re: control
P3M: I'm not overly worried about it, because like literally **everything we do at this point is tracked**. Unless we're paying with cash, every transaction we make is tracked and poured over by some sort of data analyst somewhere. Like you know, your internet history is available to those who want it. There's only so much we can keep truly secret, how off the grid we want to be, and I think that's the other [side]: how you want to **engage with society on the daily** [basis], like how much do you want to engage with it. It's almost like it's **part of the covenant we have now**, you know. There's a certain amount of ourselves that we're allowing to be available or registered.

If you look, you will find that there are more sub-topical utterances about living with smart technology here-and-now in the data that spans a variety of different questions. There might not have been any more than the first two, but it turns out there were, more than those cited above. The point to appreciate here is that if you treat them solely as responses to discrete questions, you lose the phenomenon. You lose an endogenous topic. I came to call it "the future is already here." It could have been called something different. It really doesn't matter what you call it other than it reflects in some way what an endogenous topic is about. What does matter is that naming, labelling, is an outcome of looking for sub-topical utterances that complement one another and hearably go together in speaking, in some way, to the same endogenous topic. Labelling is not a precursor to analysis, it does not drive the search for sub-topical utterances, it is not a means of pulling them together to find patterns in the data. The pull is provided by the documentary method of interpretation. Our initial sub-topical utterances – "I try and forget about it", "I've forgotten that they're doing this stuff", "It's either all or nothing I guess" are treated, for example, as the 'document of', as 'pointing to', as 'standing on behalf of' a presupposed underlying pattern: an endogenous topic about living with smart technology here-and-now. Exploiting the hearer's maxim, the individual documentary evidences (sub-topical utterances) contained in other transcripts are, in their turn, interpreted on the basis of 'what is known' about the underlying pattern: that it speaks in ways yet to be fully articulated about living with smart technology here-and-now. Each is used to elaborate the other.

Thus we come to find an endogenous topic subsequently called "the future is already here" that elaborates a discrete order of reasoning about living with smart technology here-and-now where members see themselves as augmented humans, practically cyborgs, whose daily engagement with society is governed by a new social covenant that relies on the flow of data from their smart devices. Members try to forget about it. Have forgotten that they're doing this stuff. Have given up the fight in all honesty. Because – and here's what more I found when I looked for sub-topicals that spoke about living with smart technology here-and-now – members said:

> "You can't get away." "Already today all your data anyway is shared everywhere." "There is nothing that is not shared today." "Our data's being collected a million different ways already. There's any number of different ways I think AI's being used in different services to like gather your metrics." "It's already done. They've got my heart rate, my biometrics every second of the day." "They already have access to our data totally. You start the internet and they have access to all my data. That's not an illusion, that's the truth." "Whatever we speak is being overheard by Google. That is why repeatedly those kind of news, as well as the songs, are coming, in the YouTube especially, Amazon, wherever we shop, whatever we select, that comes and comes again, and the offers whatever. So it's already AI controlled in our life, and especially the news and sport we see." "Totally our life is controlled by AI."

These are members' actual words, rough and ready as they are, and our participants had more to say about it than this, as can be found if you look in the transcripts. Nevertheless, the reader hopefully gets the gist that **their words cohere** to elaborate an **endogenous topic** that articulates a **discrete order of social reasoning** about living with smart technology here-and-now.

The collection of endogenous topics, as represented in Figure 1, is the coding frame. It reflects my interpretation of the data. It stands in need of writing up, but there is nothing special about that: you can do much as I have done above and create a narrative out of members' talk to reflect discrete orders of reasoning, as reported in our peer-reviewed study paper [24], or you can describe orders or reasoning, inserting conversational extracts to illustrate key features [e.g., 78]. In either case, the aim is to illustrate endogenous topics by describing their constitutive sub-topical features as provided **in the data**. Endogenous topic analysis (ETA) is not complicated. It doesn't require that you spend a lot of time digesting texts about interview analysis or qualitative research in order to do methodologically coherent work, at least no more than is required to defend your choice of approach (this paper hopefully provides that). Like constructive forms of analysis, ETA does take analytic skill as I have tried to **show**, but it is an **ordinary skill** that relies on mastery of natural language and being able to hear when utterances cohere and form discrete units of talk about particular things: endogenous topics.

Like constructive forms of analysis, ETA is also labour intensive and time consuming. Analysis of smart adaptive media data took several weeks, and I didn't analyse all of the data. As with most things in life, analysis is ultimately governed by time and doing enough to deliver on what needs to done within that which is available. Automated transcription helps save time, though corrections do have to be made and this may involve listening to the audio recordings to understand what was actually said, but machines nevertheless radically reduce the practical burden of transcription. As hinted at above, I didn't work through all of the transcripts, more can be found. The adequacy of analysis doesn't require it, but rather turns on whether or not analysis yields **sufficient insight to publish**. Whether it's a writing a research paper, technical report or student thesis, the adequacy of ETA doesn't turn on how exhaustive the analysis is, but on whether or not it shows us something that speaks to our research interests and / or answers our research questions. That's the measure of adequacy. Whether our analysis reveals something new, or something that speaks to and further elaborates an existing research topic; **practical adequacy**, not theoretical adequacy is what counts.

### 3.2 ETA key steps

While ETA cannot be reduced to a series of mechanistic steps – the work of identifying and elaborating endogenous topics has to be dome in each and every case – it is useful to articulate its procedural character, not least because the researcher *will* be expected to discuss their approach in any kind of publication. As with any approach, the researcher will be expected to explain what motivated the interviews, why the data was collected, what problem they sought to address, or the research questions they sought to answer. They will have to explain who their participants were, how they were recruited, and their practical relevance to the research. They will have to explain the interview format – i.e., what kind of questions were asked (structured, semi-structured, exploratory, data-driven, etc.) and why – the practicalities of data collection (i.e., where, when, how, and how much data was collected), and how the data was prepared for analysis (i.e., the mode of transcription). The researcher will also have to explain how the documentary method of interpretation was used to analyse the data and identify endogenous topics. This is how I'd sum it up. Read in conjunction with the methodological accounts and practical examples provided above, it will hopefully be instructive and useful to others.

1. Start your analysis with categories of interview question: create an analysis document, create headings that reflect the categories of question (or just the questions if you only have a few, or if particular questions are very pronounced and people had a lot to say about them), then map responses under the headings.

2. Map responses by cutting extracts of talk from transcripts and pasting them under the appropriate question headings in the analysis document, then highlight what is topically interesting in the talk so you can quickly refer back to it and see if other utterances complement it.
3. As you read through transcripts, look out for sub-topical utterances that speak in various ways about topics already identified, and those that introduce and articulate new topics.
4. Collect together sub-topical utterances that hearably go together, that cohere around and elaborate a topic, and give the topic a working title or name to reflect what it is about.
5. Don't be constrained by the questions, topical talk can also be used to make jumps to new topics of interest to interviewees, so follow topics and map them on their own if they do not map to interview questions.
6. Use the hearer's maxim and documentary method to deal with contingencies and choices: ask are 'these' utterances talking about the same sort of thing as 'those' utterances, or are they talking about something else?

The result of your analysis will be a coding frame that elaborates categories of question. Some questions may be quite one-dimensional, i.e., characterised by topical talk that follows much the same line of reasoning. However, you should expect most questions to be elaborated by a diverse array of endogenous topics, each populated with a collection of conversational extracts consisting of sub-topical utterances that articulate members' reasoning. No set number of sub-topicals will elaborate an endogenous topic, some topics will collect together many extracts, others only a few, even only one in some cases. It all depends on the data, on what people said. The coding frame will also include endogenous topics that have nothing to do with the questions asked, but which emerge in participants discussions. The coding frame is your interpretation of the data. It organises the many thousands of words contained in transcripts, or at least a subset extracted from them, into meaningful patterns that articulate members' reasoning about issues of relevance to your research. You can write them up as best suits the kind of research you are doing, narratively, illustratively, or indeed in any way that works for you and the recipients of your research.

### 3.3 Limitations of ETA

As noted when discussing the notion of member in section 2, language cannot simply be reduced to words. Language is culture. Not culture as in art, and certainly not culture as in an object to be theorised [35], but culture as an "embedded phenomenon in language-in-use" [47]. As Garfinkel and Sacks told us, in speaking a natural language, members can "be heard to be engaged in the objective production and objective display of commonsense knowledge of everyday activities as observable and reportable phenomenon" [32]. This means that the "definite sense" of any utterance is "indexical" to the actual "circumstances of its use" (ibid.), i.e., the everyday activities in which it is culturally embedded. This simple fact about language is enormously consequential for the conduct and analysis of interviews, insofar as the actual circumstances of use are **not** contained **within** the interview. Commonsense knowledge is not an equally distributed stock of knowledge about everyday activities. Many of us know how to catch a train or plane and read a newspaper, for example, but few know how trains, planes and print production are organised. As ethnography shows us, we can learn [e.g., 38, 41, 15], but commonsense knowledge of train coordination, air traffic control, or print production is not necessarily available in an interview. It's not that we can't ask questions about things we don't know, obviously we can, it's that we won't necessarily understand the response, we won't "know how to hear it" [32].

Wittgenstein once said, "the limits of my language means the limits of my world" [82]. The culturally embedded language and indexical expressions people use to describe and reason about their everyday activities, may not always be intelligible to us. Their sense turns on 'vulgar competence' [34], i.e., on the competence of the practitioner. That's not always available to the analyst. We may have mastery of natural language but not possess vulgar competence in the activities being discussed (e.g., controlling aircraft or trains), hence our not being able to know how to hear practitioners' talk. Nevertheless, a surprising array of talk in domains unfamiliar to the analyst is intelligible. Take, for example, the 'pulsar' paper [33]. It presents an analysis of a tape recording of the conversations between several astronomers at the Steward Observatory in Arizona on the evening of January 16, 1969, during which they discovered a rotating neutron star emitting beams of electromagnetic radiation some 6,500 light years away from Earth in the Crab Nebula. Garfinkel and his colleagues were not astronomers and had no training in astronomy, yet their analysis reveals how the discovery was made in details of the astronomers' "shop work and shop talk." It turns out that astronomers speak ordinary language too. The same applies to all manner of professional activities, including scientific practice. Discussions of professional practice may well be opaque and even unintelligible on occasion, but not necessarily so, and if we are sensitive to the limits of our language in the course of conducting interviews it may well be possible to elicit and elaborate endogenous topics that usefully inform our research even in highly specialised domains [see 31].

## 4. Established methods for analysing interviews

The reader might ask, quite rightly, how is ETA different from established approaches? To answer that question we first need to understand how established approaches to analysing interviews work. Methods for analysing interviews fall into one of two basic camps: 'top down' deductive approaches, which employ pre-defined categories that reflect, for example, the key concepts of a theory or hypothesis and subsequently map patterns in the interview data to them; and inductive approaches, where patterns emerge 'bottom up' from examination of the data; both approaches may be employed in practice. Methods for analysing interviews might also be said to fall into two basic camps: quantitative or qualitative, though again, in practice, methods may be 'mixed'. Mixed methods approaches are not considered here. Neither are quantitative methods, other than to say they seek to collect responses from a large, representative sample of respondents and find statistically valid patterns within them. They are supported by dedicated software packages such as SPSS. Quantitative approaches are very different to qualitative approaches, which replace interest in statistical patterns with understanding the human experience, behaviour and reasoning that underpins and gives rise to such patterns [22]. Qualitative approaches are also supported by dedicated software packages, such as NVivo, Transana and ATLAS.ti. These are not considered here either [see 16], other than to say they offer varying levels of support for approaches to interview analysis that turn on coding utterances in transcripts to find patterns within the many thousands of words they may contain. These approaches include content analysis, grounded theory, interpretative phenomenological analysis, and thematic analysis, all commonly used in HCI. Other qualitative approaches to interview analysis are available – e.g., hermeneutics, narrative analysis, discourse analysis [see 26] – but the focus here is on code-based approaches as these are more commonplace in HCI and have *prima facie* resonance with ETA.

### 4.1 Content analysis

Content analysis (CA) originated in the late 19th Century and was initially a quantitative approach to media analysis [74]. A qualitative version emerged in the 1950s [44]. It is often used in HCI today to analyse social media [52, 73]. It has also been used to analyse other texts, including game reviews [60, 61] and codes of conduct in multiplayer games [37], research papers on automatic speech recognition [63], transparency and openness guidelines in HCI journals [3], as well as interviews [25, 43, 79]. Qualitative CA stems from Siegfried Kracauer's reflections on the 'black sheep' of the family:

> " … quantitative analysis is more 'impressionistic' than its champions are inclined to admit. All of them, incidentally, readily grant the need for qualitative reasoning in the initial stages of category formation. They more rarely admit, however, that the quantification processes themselves often require much conjecturing which is not in actuality tied to objective, impersonal definitions. A recent quantitative analysis of Voice of America and BBC broadcasts classifies the 'style' of contextual units as 'matter-of-fact', 'mildly emotional' and 'highly emotional' … value-laden terms … The example is not unique. Numerous quantitative analyses are similarly threaded with impressionistic judgments. And these judgments may in fact be more unaccountable than those found in communications studies of a predominantly qualitative nature. For within the framework of quantitative analysis, qualitative exegesis is condemned to playing a black sheep role." [44]

Kracauer didn't see qualitative CA as being radically different to quantitative CA. Rather, in drawing the distinction, he wanted researchers to recognize that quantitative analysis both originate and culminate in qualitative considerations at the heart of which is "the selection and rational organization of such categories as condense the substantive meanings of the given text" (ibid.). The coding frame in other words.

The coding frame is the outcome of qualitative interpretation of the data. It consists of a collection of categories built up through careful inspection and examination of the data, that articulate discernible patterns within it. Coding frames are constructed in steps or stages, which vary depending on approach. In CA, a representative sample of the data must first be selected to drive initial construction. Main categories must be identified to structure the coding frame and sub-categories generated to elaborate them. Main categories should be 'unidimensional' and apply to only one aspect of the data, sub-categories should be 'mutually exclusive' and apply to only one main category. The coding frame as a whole will be multi-dimensional, and there may be relationships between categories, but the categories that make up the coding frame are essentially distinct. Development of the coding frame will "**in part be data-driven**" to ensure a good fit with the material, but categories will also be "generated conceptually" [74]. As Schreier (ibid.) puts it,

> " … when defining the categories, one will usually go beyond the specifics of any particular passage. Instead, the meaning of the passage will be taken to a **higher level of abstraction**, resulting in categories that apply to a number … different passages."

Categories are then defined, which entails providing an appropriate name or label that describes what a category refers to; specifying its characteristic features, including 'indicators' to support recognition of phenomena in the data and examples that illustrate the category. Optionally, decision rules might also be formulated to help coders deal with uncertainty. The coding frame is then expanded to include further data, and revised if necessary through further category development, prior to segmentation or assessment of parts of the coding frame, often by multiple independent coders. The coding frame as a whole is then piloted, again often by multiple independent coders, in order to validate the consistency with which it can be applied. This includes assessing the adequacy of category definition, the need for decision rules, and determining the degree of consistency by calculating a coefficient of agreement. Following final modifications, the coding frame may then be applied to all remaining data. The presentation of findings involves presenting the frame and illustrating it through quotes.

### 4.2 Grounded theory

While the interpretation of data through the construction of coding frames lies at the heart of many approaches to interview analysis, quantification *does not*, as can be seen if we turn to Grounded Theory, one of the most influential qualitative approaches in the social sciences to date (much more so than CA). Grounded Theory (GT) is a sociological approach "discovered" in the 1960s by Barney Glaser and Anselm Strauss [36]. GT is often used to analyse interviews in HCI. Some recent examples include using it to understand frustration [4], confusion [18] and emotional experiences in gameplay [19], design challenges confronting migrant women [83], online responses to hate speech [58], the expansion of spiritual care into online spaces [5], and the role of technology in enabling climate resilience [39]. GT replaces quantitative sampling with **theoretical sampling**, an iterative process for generating theory that turns on coding.

> "Theoretical sampling is done in order to discover categories and their properties, and to suggest their interrelationships into a theory. Statistical sampling is done to obtain accurate evidence on distributions of people among categories to be used in descriptions or verifications. Thus, in each type of research the 'adequate sample' that we should look for (as researchers and readers of research) is very different." [36]

The adequacy of theoretical sampling does not turn on issues to do with sample size, representativeness, multi- and inter-coder reliability, calculations of coefficient agreement, etc. Contrary to common misunderstandings in HCI about grounded theory [see 54], GT is "not designed (as methods of quantitative analysis are) to guarantee that two analysts working independently with the same data will achieve the same result" [36]. Instead, the adequacy of theoretic sampling turns on the analysts' ability to develop a substantive theory about the study topic (e.g., patient care or gang behaviour) that is grounded in the data. GT is an inductive method of theory development that turns on iterative data collection (no representative sample here) and the constant comparative method, an analytic process that first entails comparing incidents applicable to each category in order to develop the coding frame. The word 'incident' should be read broadly as per dictionary definition and thus be treated as referring to an instance of something happening, an event or occurrence of a particular kind. The aim is to "code each incident in the data into as many categories of analysis as possible, as categories emerge or as data emerge that fit an existing category". The comparative character of this method lies in comparing the incident to hand to "with others previously coded in the same category" to determine its fit. Thus, the "constant comparison of incidents very soon starts to generate theoretical properties of the category."

A category is a conceptual element of the emerging theory (a main code as it were), a property a conceptual aspect or element of a category. Importantly, both categories and properties are **"concepts indicated by the data"**, they are **"not the data itself"**. They are either "abstracted from the language of the research situation" (i.e., from participants' actual talk) and otherwise "constructed" by the analyst in the process of comparing incidents. Either way, it is important that the analyst occasionally "stop coding and record a memo" to explain how the abstractions and analytic constructs enabling the specification of categories and their properties are "grounded in the data, not speculative conclusions." Data should be coded only enough to generate theory, with enough being determined when "saturation" is reached and no additional data are found to further elaborate the properties of the discovered categories. Then the analyst can move to the integration of categories and their properties. Here the focus shifts "from comparison of incident with incident, to comparison of incident with properties", such that "diverse properties themselves start to become integrated." As modifications become fewer and fewer, the analyst "delimits" the theory. The theory, in short, solidifies, and modifications shift to reducing the original set of categories or their properties to enable the analyst to "formulate the theory with a smaller set of higher level concepts." The coding frame constitutes an "analytic framework" that "forms a systematic substantive theory". The theory stands in need of writing up, about which Glaser and Strauss have very little to say apart from "it is first necessary to collate the memos on each category" and "one can return to the coded data when necessary to validate a suggested point, pinpoint data and provide illustrations" [36].

### 4.3 Interpretative phenomenological analysis

Interpretative phenomenological analysis (IPA) offers an approach that is similar to and but different from grounded theory [75]. It has its origins in psychological research and is based on first-person accounts, usually semi-structured interviews and this how IPA is typically used in HCI. It has recently been used to understand how the experiences of physically disabled players are reflected in guidelines for movement-based games [53], mental workload [55], synaesthesia [80], personal informatics [1], and embodiment in social VR [45]. IPA involves case-by-case analysis of small, homogeneous groups (again, no large, representative sample here), and examines similarities and differences within them by identifying what *matters* to participants and exploring what these things *mean* to participants [46].

> " … meaning is central, and the aim is to try to understand the content and complexity of those meanings rather than measure their frequency. This involves the investigator engaging in an interpretative relationship with the transcript. While one is attempting to capture and do justice to the meanings of the respondents to learn about their mental and social world, those **meanings are not transparently available** – they must be obtained through a sustained engagement with the text and a process of interpretation." [75]

The process of engagement and interpretation first involves immersion in the data, reading individual transcripts and noting initial observations, which may be about the interview, the transcript or the analyst's initial thoughts and preconceptions [62]. Next comes phenomenological coding, which involves line-by-line reading of a new (clean) transcript and annotation of "'objects of concern' (anything that matters to the participants) and 'experiential claims' (linguistic and narrative clues as to the meaning of those objects)" [46]. These annotations are typically made in the left-hand margin of a transcript and reflect a free or open textual analysis. There are no rules about what is commented upon, and there is no requirement, as Smith and Osborn [75] tell us, to divide the text into meaning units and assign a comment for each unit. Having thus mapped out the "phenomenological core" of the data, analytic attention then turns to identifying emergent themes at the level of individual cases. This is driven by the analyst's annotations and clustering associated objects of concern and experiential claims into small units of meaning, i.e., "bundles" of terms or phrases that reflect conceptual similarities and tentatively capture emerging ideas, to which the analyst can attach a thematic label. As Smith and Osborn (ibid.) put it, "the skill at this stage is finding expressions which are high level enough to allow theoretical connections within and across cases but which are still grounded in the particularity of the specific thing said." The process is repeated with each participant and a table of themes is constructed for each transcript [76].

The table lists emergent themes and includes a relevant short extract from the transcript for each, followed by the line number, so that it is possible to return to the transcript and check the data (ibid). While themes are rooted as such in the data they are, nevertheless, "interpretative" and thus **"step beyond" the data** [46]. Interpretive coding follows, and entails looking for analytic connections between emerging themes in order to explicitly consider the interpretative ideas emerging from the analyst's engagement with the data. Interpretive coding is construed as a dialogue between the analyst and the coded data, which aims to produce an "interpretative synthesis of their analytic work" [62]. The synthesis is done by **clustering themes together to reflect their conceptual similarities** and analytical or theoretical ordering. Some of the themes may be dropped at this stage, if they do not fit well with the emerging structure or because they have a weak evidential base. Some may emerge as "superordinate concepts" – Smith and Osborn [75] tell us to imagine a magnet; some themes thus pull others in and help make sense of them. Either way, interpretative coding should develop from, and connect to, the phenomenological core. It is therefore important to check the transcripts to make sure the interpretative connections being drawn between themes actually exist, a process that may be supported by compiling directories of participant's phrases related to themes (ibid.). Nonetheless, as Larkin and Thompson [46] tell us, the interpretative synthesis of the analyst's work is not constrained by the data. Indeed, it is important that the analyst "**open a dialogue with theory**" in order to see the phenomenon from a perspective that can lead to a "more insightful account." The subsequent collection of themes should be formally structured in a coherently ordered table. The table is a heuristic device that provides a graphic representation of the patterns of meaning found in the whole data set and illustrates the relationships between themes therein (ibid.). It displays superordinate and associated themes, and includes an identifier that indicates where instances of each theme can be found [75]. The analysts may then turn to writing up and translating the themes into a narrative account. This entails writing up themes one by one. Each theme needs to be introduced and then illustrated with extracts from the participants, which are in turn followed by analytic comments from the authors [76]. The narrative account may span different levels of interpretation, including low-level interpretation of the data and a high-level interpretation that generates new analytic insights [62].

### 4.4 Thematic analysis

Thematic analysis (TA) also emerged out of psychological research [8]. TA has been described as "a staple of qualitative HCI research" [13]. Bowman et al. [7] identified 78 CHI papers that used TA in the area of healthcare alone between 2012 and 2021. Recent examples include using TA to identify dark patterns in social networking services [56], themes to create and support gender euphoria in video games [48], and (I quote) themes to design "shitty user experiences" [20], as well as TA being used to analyse interviews and understand problematic interactions with conversational UIs [57], how users interpret visualisations [14], and non-computer scientists experiences of coding with computers [59] *amongst many others*. Thematic analysis is by far and away the most popular method of qualitive analysis in HCI today, not that its use is necessarily approved of by its authors. Braun and Clarke, for example, the most cited of TA's authors inside and outside of HCI, say:

> "Our approach is often being used in … applied research … It seems … [it] is simply applied to the data, akin to a quantitative method or tool.
>
> Far too often we encounter 'I followed Braun and Clarke's six steps', and then they list the phases … That reflects a … positivist approach – you lay out the steps, rather than discussing your practice.
>
> Then they will write 'we did consensus coding, we measured inter-rater reliability, we used a codebook.' Where is that in our paper?

Saturation and coding reliability measures. These practices come from a very different position from our approach.

Ironically, the best outcome would be for fewer people to use our approach, because it's not the approach they need, they need something else.

*Extracts from Virginia Braun and Victoria Clarke in conversation* [12].

As noted above, Braun and Clarke's approach consists of six steps [8]. TA thus begins with data familiarization, reading transcripts and noting down initial ideas. Initial codes are then generated across the data set, either through inductive or deductive engagement with the data or both, to "organise the data into meaningful groups." The search for themes may then begin, which is "where interpretative analysis of the data occurs." Interpretive analysis entails looking to see "how different codes may combine" to create themes and "collating all the relevant coded data extracts" to elaborate them. Themes are then reviewed to check that they "work in relation to the coded extracts", the data is reworked if necessary and further themes developed, and a "thematic map" that shows how themes fit together is generated. The analyst may then progress to defining and naming themes, including further refining themes by "identifying the 'essence' of what each theme is about" and "determining what aspect of the data each theme captures", prior to producing a report, a step that involves the selection of "vivid, compelling examples" with which to elaborate themes (ibid.).

> "Themes are analytic outputs developed through and from the creative labour of our coding. They reflect considerable analytic 'work,' and are actively created by the researcher at the intersection of data, analytic process and subjectivity. **Themes … are not 'in' the data**, waiting to be identified and retrieved by the researcher. **Themes are creative and interpretive stories about the data**, produced at the intersection of the researcher's theoretical assumptions, their analytic resources and skill, and the data themselves." [9]

TA is often seen and treated as being an atheoretical method – a view underpinned by Braun and Clarke's original description of TA as being independent of theory thus rendering the approach more accessible and flexible [8] – but as the author's explain "that's probably the most fundamental misunderstanding of our approach" [12]. The term "reflexivity" was added to subsequent writings to reflect the importance of theory to TA. Not explanatory theories of human behaviour the authors stress, but "bigger ideas" to do with ontology and epistemology in qualitative research that "provide TA with analytic power and analytic validity" [10] enabling a "methodologically coherent TA" and "knowing TA researcher" [11].

**4.5 How established qualitative methods for analysing interviews work**

The above descriptions of how established approaches work are summarised below in Table 1 and allow us to explain what we mean when we say they are constructive analytic approaches and that ETA is not. It is obvious (plain to see) when you look at Table 1 that ETA is quite different from the established approaches considered here. You can see at a glance that the work of coding is significantly *reduced*, limited to collating and labelling endogenous topics and sub-topics. ETA only 'codes' endogenous topics and sub-topics. Nevertheless, one might think that endogenous topics and sub-topics are just the same as main categories and sub-categories, or categories and properties, or clusters of annotations collected into units of meaning, or data organised into meaningful groups combined to identify themes. That endogenous topics and sub-topics are equivalent. Just different words for the same things.

To treat them as equivalent would be to reduce language to mere words. To ignore the **indexical** character of words and the actual circumstances of their use. One would have be situationally disengaged from the work practices in which these words are embedded, and thus fail to see what is unique about them, to assume that they are all essentially the same. Main categories and sub-categories in content analysis are "in part" identified through the data, for example, whereas categories and properties in grounded theory are "not in the data" at all, they are concepts "indicated" by it. One could try and turn it round and say categories and sub-categories are much the same as categories and properties insofar as they are conceptually generated, but then that would lose the data-driven work that is essential to the identification of categories and sub-categories or indication of categories and properties. IPA and TA appear to have more in common. Annotations made during phenomenological coding are clustered together with bundles of data to identify themes in IPA, just as initial codes that organise data into meaningful groups are subsequently combined to create themes in TA. But this *single* similarity does not make their working practices the same. It is plain to see in Table 1 that they are not. There is no equivalence in the work of coding amongst established approaches. It gets done in different ways and the different words used to describe the doing reflect this working fact.

| **Content Analysis** | **Grounded Theory** | **IPA** | **Thematic Analysis** | **ETA** |
|---|---|---|---|---|
| Data (transcripts) | Iterative data collection | Data (transcripts) | Data (transcripts) | Data (transcripts) |
| Select representative data for coding | Data (transcripts) | Immersion in data: read transcript(s) and note initial observations | Data familiarisation: read transcript(s) and note initial observations | Read each transcript and identify endogenous topics:<br>a) by collating **sub-topical utterances** that cohere around interview questions; label to reflect topical coherence<br>b) by collating **sub-topical utterances** that cohere around themselves; label to reflect topical coherence<br>Collate in working document that reflects endogenous topics, sub-topics, and associated utterances |
| Identify main categories and sub-categories: **in part data-driven**, but also generated conceptually | Identify categories and their properties: by comparing 'incidents' in the data - categories and properties are concepts 'indicated' by the data **'not in the data'**<br>Data is coded to saturation point, when no new incidents found | Phenomenological coding: **annotations** marking out 'objects of concern' and 'experiential claims' in a clean transcript | Initial coding: **organise data into meaningful groups** | Write up findings narratively or on a case by case basis; either way, each endogenous topic is illustrated by describing its sub-topical features as provided **in the data** |
| Define categories at a **higher level of abstraction** than particular passages in transcripts; specify indicators and (optionally) decision rules to enable application to data | Delimit theory: **reduce** original set of categories **to a smaller set of higher level concepts** | Identify emergent themes: **cluster annotations** into units of meaning and **attach a thematic label** | Interpretative analysis: **look to see how different codes combine to create themes** and collate all relevant coded data extracts to elaborate themes | |
| Independent coders assess and validate coding frame on new data | Write up theory: supported by memos and illustrations from the data | Interpretative coding: **cluster themes to reflect conceptual similarities and analytic connections** | **Develop a thematic map**: show how themes fit together | |
| Coding frame piloted and deployed, quotes used to illustrate findings | | **Structure themes** in a table to display superordinate and associated themes: each supported by narrative account illustrated with data extracts | **Define and name themes**: identify the essence of what each theme is about; write up with vivid examples using data extracts | |

**Table 1.** Comparison of established approaches and ETA.

However, when we look at the work these words speak of, when we consider their indexical character, a common practice is discernible. Coding in content analysis is, as noted, only "**in part data-driven**" and turns on a "**higher level of abstraction**" [74]. Coding in grounded theory entails developing **"concepts indicated by the data"**, concepts "**abstracted from**" the data and otherwise "**constructed**" by the analyst in the process of comparing incidents [36]. Coding in IPA **"steps beyond"** the data, **clustering themes together to reflect conceptual similarities** [62]. TA organises codes into "**themes not 'in' the data**" to enable the analyst to develop "**creative and interpretive stories about the data**" [11]. Baccus [2] calls this "real world theorising" and notes that it produces "constructive analytic accounts formulated by appropriating real-world talk." Central to the production is a general method of **interpretation through abstraction**, of stepping beyond the data, of moving to a higher level of abstraction indicated by the data, of constructing categories, concepts and themes from the data that are **not in the data**.

Abstraction is seen as necessary, unavoidable even, for several interrelated reasons. Contrary to the members' perspective, it is received wisdom in the social sciences that the meaning of talk is **opaque**. Smith and Osborn have already told us that **"**meanings are not transparently available" and must be arrived through "a process of interpretation" [75]. Braun and Clarke similarly tell us that "the meanings of the data aren't obvious" [12]. As Roulston puts in describing how to go about analysing interviews, we should not "assume that a person's words are a transparent window – they're more like the smoky, veiled, dirty window that you're trying to see through" [65]. Consequently, it is also received wisdom that interpretation is necessarily **theory driven**. As Braun and Clarke [11] put it,

> "Any analytic method contains theoretically embedded assumptions, whether acknowledged or not, when applied to the analysis of a particular dataset. This means TA research, indeed **all** research, cannot be conducted in a theoretical vacuum. Even if theory isn't acknowledged, it's there, lurking, to bolster and / or undermine the validity of what you claim."

Acknowledging the inevitability of theorising obliges the researcher to interpret the data and **index** or connect it to a contingent array of academic literatures as they go about analysis [51]. Thus, constructive analysis proceeds through a general method of abstraction.

Doesn't ETA do just the same? Isn't labelling a collection of utterances – *the future is already here*, for example – essentially an interpretive act of abstraction? Doesn't the label apply to a number of different passages? Isn't it a concept indicated by those passages? A discrete unit of meaning that combines different coded utterances to create a theme that is not in the data? An interpretive story about the data produced at the intersection of the researcher's theoretical assumptions, their analytic resources and skill, and the data themselves?

It's a beguiling proposition on the face of it from a situationally disengaged point of view. The reality is that we are dealing with an entirely different working practice. ETA doesn't suspend membership competence and our mastery of natural language. It doesn't assume at the outset that the meaning of talk is opaque and thus treat it as something that *necessarily requires* we **step beyond the data** and interpret it through constructive analytic practices. An endogenous topic doesn't exist at higher level of abstraction. The label is a characterisation **of** the data, not a story **about** the data that is **not in** the data. Remember, an endogenous topic is **no more than** a collection of sub-topical utterances, labelled to reflect the topical coherence of talk. Constructive analytic accounts do not attend to the topical coherence of talk (it is opaque by default) and require of the analyst much **more than** labelling: e.g., the specification of indicators and decision rules to provide for the application of higher level categories to the data; delimitation practices that reduce categories and their properties to higher level concepts; clustering of themes to reflect conceptual similarities and analytic connections; structuring of themes to display superordinate relationships; the definition and naming of themes after the fact; etc. Ultimately, endogenous topics and their constitutive sub-topics are in the data, oriented objects **in the talk**, which is why we say they are endogenous. They are available to members through the documentary method of interpretation and concomitant use of the hearers' maxim. They are not available to constructive analytic practices that abstract from the data to make higher-level connections and open a dialogue with theory. We can always do that afterwards, should we want to, but first we should explicate the real-world character of talk. It is not available to a general method of interpretation that eschews membership competence or situationally disengaged efforts to render ETA equivalent.

### 4.6 The unacknowledged equivalence between established approaches and ETA

That said, there is an equivalence between established approaches and ETA that touches on the general method of interpretation by abstraction, but not as conjectured. The DMI is a mundane method of fact-finding, a members' method that turns on "commonsense situations of choice" [27], e.g., deciding when one series of utterances are speaking about the same sort of thing as another, or speaking about something new, or something different. Commonsense situations of choice are resolved in ETA through the use of the documentary method of interpretation and hearer's maxim, which allows us to collect specific utterances together to detect distinctive patterns in the data (endogenous topics). Commonsense situations of choice are resolved by the constructive analyst in distinctive procedural ways outlined above and listed in Table 1, e.g., by formulating higher level abstractions, comparing incidents, clustering themes and combining codes, formulating decision rules, etc. These procedures necessarily turn on the analyst's ability to determine the **correspondence between an actual appearance** (some piece of talk) **and a presupposed underlying pattern** (a category, concept or theme). In other words, the constructive analyst must *also* use the documentary method of interpretation to fashion a relationship between actual utterances in a transcript and higher-level categories, concepts or themes that are not in the data. The equivalence between ETA and established approaches is not that ETA is a form of constructive analysis that trades on the general method of interpretation by abstraction then, but that established approaches trade on the DMI. This isn't to say that established approaches use ETA – ETA is our configuration of the DMI that relies on the hearer's maxim – but that they must treat utterances as documents of, as pointing to, as speaking on behalf of some presupposed underlying pattern, where that pattern is not an endogenous topic but a higher-level category, concept or theme. **Thus, the equivalence lies in the use of the DMI**.

## 5. Dispensing with constructive analytic practice and real world theorising

One of the reviewers of this paper asked is there a need for the DMI? Generally speaking it would appear that there is. As Garfinkel showed us, the DMI is a ubiquitous method of fact-finding in a lay and professional studies of all kinds that seek to find patterns in human behaviour, enabled through local or situated configurations and practices. The analysis of interviews is no different. Yet it is unacknowledged by established approaches, a **tacit** feature of their conduct. Ethnomethodological studies have shown that scientific practice trades on members' methods [e.g., 33, 50, 49, 31]. The qualitative analysis of interviews is no exception. It is time the DMI was acknowledged and its use legitimated in professional practice. Doing so presents us with a choice: it is not possible to disconnect the DMI from constructive analytic practice – commonsense situations of choice have to be dealt with even by social scientists and HCI researchers – but it is possible to use the DMI **on its own** and dispense with real world theorising. As Lynch reminds us,

> "Such remarks may suggest an unquestioned faith in the world's capacity to speak for itself … these days anyone who … propos[es] that the real world, and not a literary tradition, is the primary source of ideas is liable to provoke knowing winks and muffled laughter among the cognoscenti … It is no less risky to assert … that transcriptions, collections of transcribed excerpts, and written analytic descriptions are transparent media through which naturally occurring interactions [including mundane reasoning] are represented. It can be difficult indeed … to express a preference for deriving ideas from the world, without seeming naive … [However] practical experience of the 'real world' has a leading role in ethnomethodology … the 'real world' is the everyday life-world and not the analytical world of a science … … … practitioners and their activities not only furnish information for social scientists to interpret, they provide tutorials about relevant topics." [51]

Dispensing with real world theorising allows the analyst to reveal the endogenous topical organisation of member's talk across a corpus of transcripts. ETA teaches us what is important to members of the cohort, provides tutorials about relevant topics oriented to and elaborated in their talk, and in doing so articulates discrete orders of mundane reasoning that are not reducible to individuals. One could, of course, separate out particular sub-topical utterances and trace them back to particular participants, but that would lose the phenomenon: how particular sub-topical utterances made by particular participants cohere and thus contribute to a much larger **social dialogue** in which interviewer and interviewees are engaged. The participants might not all be speaking at the same time, but they are contributing to a broad discussion nevertheless. That's what a collection of transcripts is, a broad discussion of some issues of interest to the researcher / interviewer, contributed to and elaborated by multiple parties. Dispensing with real world theorising allows us to see and treat the discussion for what it is then: a socially produced, multi-party conversation that articulates patterns of **social reasoning** (endogenous topics) addressing issues of relevance to our research. The DMI thus invites us to respect the integrity of the interview as a social phenomenon possessed of its own naturally occurring means of accountability. It offers a unique social perspective on interviews and analytic practice for their interpretation that we call endogenous topic analysis (ETA) to reflect its distinctive contribution.

Our configuration of the DMI is a disciplined effort to set theorising aside and respect the integrity of the interview as a naturally accountable, social phenomenon. An effort to be **indifferent** to theorising and the demand that we must necessarily make our interpretation accountable to it. Language isn't opaque by default. Theorising isn't unavoidable, it is an **option**. One way of doing the job, but not the only way. There is an ethnomethodological alternative.

> "Ethnomethodology's attitude is prescribed under the rubric of 'ethnomethodological indifference' … There is nothing heroic about indifference … it is a matter of explicating situations with a full attention to their ordinary accountability … not taking up a gratuitous 'scientific' instrument: a social science model, method, or scheme of rationality for observing, analyzing, and evaluating **what members already can see and describe as a matter of course**. (ibid.)

The ordinary accountability of talk, its natural accountability to members, to masters of natural language, underwrites the credibility of ETA as alternate method of social analysis commensurate with ethnomethodology's methods and policies [34]. Chief among them is the **unique adequacy requirement of methods**: that the analyst be in a concerted competence of methods with members (ibid.). ETA is built on membership competence – mastery of natural language – which it uses with the DMI to deliver insight into members' social reasoning about topics of relevance to research.

It might be asked, not unreasonably, how we might assess, in any particular case, the rigour and quality of ETA? How do we know the results are trustworthy and reliable? Like Glaser and Straus, we have no interest in inter-coder or inter-rater reliability, coefficients of agreement, representative samples, etc. It may be fashionable to try and make qualitative research accountable to quantification, but it is not a methodological feature or requirement of a great many established qualitative approaches. As McDonald et al. [54] put in their review of inter-rater reliability (IRR) in qualitative research in HCI and CSCW,

> " … researchers have sometimes made the mistake of evaluating qualitative research reports using the standards of quantitative research, expecting IRR regardless of the nature of the qualitative research … for many methods, reliability measures and IRR don't make sense. They may even be outright harmful."
>
> " … most qualitative CSCW and HCI papers code data as part of their research process and most provide some description of the method they use to do so. Far fewer report the number of coders involved in the process and even fewer use IRR."
>
> "In sum: there is no one correct way to approach reliability in qualitative research. Different epistemologies invite different frames."

We have described the 'framing' of ETA above and how it provides a unique ethnomethodological orientation to and analysis of the topical organisation of talk through the use of membership competence. How do we know in any particular case that ETA is sound? In short, its results should be available to membership competence. Members should be able to see endogenous topics in the data. As Garfinkel puts it, *"If it's not in the look of things, then where in the world do you think you're going to find it"* [30], a point similarly made by Harvey Sacks when describing the results of his analyses and saying, *"anybody can go and look, and see that the thing seems to be as we said"* [68]. That is the virtue of membership competence. It takes no special training to see if things look like the analyst said they do, only mastery of natural language, which most people possess, even HCI researchers. More generally, and as recently expounded at length [22, 23], particular studies in endogenous topic analysis should, as any form of qualitative analysis, be assessed through the use of appropriate quality criteria, including the intelligibility of methodological approach; the apodicity (visibility) of findings; the introduction, extension or refinement of sensitizing concepts; the potential utility of the analytic insights provided; and their relevance to HCI. Our example study already meets these criteria [24].

## 6. Conclusion

Interviews are commonplace in HCI, routinely used to understand users and support evaluation of interactive systems. We have reviewed how established approaches to the analysis of interviews work through constructive 'coding' practices that rely on a general method of interpretation through abstraction to produce higher-level categories, concepts and themes that are **not in the data**, but **indicated by the data**, tell stories **about the data**, and thus allow the analyst to open a dialogue with theory. This way of working (real world theorising) trades on the fundamental assumption that the meaning of talk is essentially opaque and thus requires the use of specialised interpretive practices. It contrasts with the ethnomethodological alternative: endogenous topic analysis. ETA proceeds on the basis of membership competence. A member is a master of natural language, an ordinary language speaker. The fundamental assumption of established approaches does not hold for ETA. Members understand the meaning of talk, being able to do so is a necessary feature of their mastery of natural language. ETA provides a unique orientation to the analysis of interviews that requires no special training. It allows the analyst to make a unique contribution and elaborate the **topical organisation of talk** that is contained **in the data** (transcripts of real-world talk) and thus make visible discrete orders of social reasoning relevant to research. We have outlined the ethnomethodological foundations of ETA, including the use of the documentary method of interpretation and hearer's maxim to identify endogenous topics, and provided transcripts and a tutorial to support others in the use of ETA. In keeping with ethnomethodological policies and methods, the methodological account and tutorial should be treated as a lebenswelt pair furnishing the situationally engaged reader with **instructions** for doing a real world job.

### Disclosure statement

No potential conflict of interest was reported by the author(s).

### Data availability statement

Data supporting the results or analyses presented in the paper can be found here: http://doi.org/10.17639/nott.7408

### Funding

This work was supported by the Engineering and Physical Sciences Research Council grant number EP/S02767X/1. The human studies reported in this work were approved by the Faculty of Arts and Social Sciences and Lancaster University Management School (FASSLUMS-2022-0762-RECR-3) and School of Computer Science, University of Nottingham (CS-2022-R17).